\documentclass[twocolumn]{revtex4}
\usepackage{epsfig}
\newcommand{\m}{\mu}

\def\equ#1{(\ref{#1})}

\begin{document}

\title{A note on black hole entropy, area spectrum, and evaporation}

\author{C.A.S.Silva} \email{calex@fisica.ufc.br}
\author{R.R. Landim}
\email{renan@fisica.ufc.br
}

\affiliation{Departamento de F\'{\i}sica, Universidade Federal do
Cear\'a - Caixa Postal 6030, CEP 60455-760, Fortaleza, Cear\'a, Brazil}

\begin{abstract}
\noindent We argue that a process where a fuzzy space splits in two
others can be used to explain the origin of the black hole entropy,
and why a ``generalized second law of thermodynamics''
appears to hold in the presence of black holes. We reach the Bekenstein-Hawking
formula from the count of the microstates of a black hole modeled
by a fuzzy space. In this approach, a discrete area spectrum for the
black hole, which becomes increasingly spaced as the black hole approaches
the Planck scale, is obtained. We show that, as a consequence of this,
the black hole radiation becomes less and less entropic as the black
hole evaporates, in a way that some information about its initial state
could be recovered. 
\end{abstract}

\maketitle


\indent It is not an exaggeration to say that one of the most exciting
predictions of general relativity is that there may exist black holes.
This is mainly due to the believe that black holes may play a major
role in our attempts to shed some light on the nature of a quantum
theory of gravity such as the role played by atoms in the early development
of quantum mechanics.

Recent results have shown that black hole have thermodynamics
properties like entropy and temperature, and as a consequence of the
instability of the vacuum in strong gravitational fields, black holes
are sources of quantum radiation \cite{sw.hawking-cmp43,sw.hawking-prd14,jd.bekenstein-lnc4}.

Some time after, string theory and loop quantum gravity, argued that
the origin of the black hole entropy must be related with the quantum structure of the spacetime \cite{a.strominger-plb379,A. Ashtekar-prl80}.
However the computation of black hole entropy in the semiclassical
and furthermore in quantum regime has been a very difficult, and still
unsolved problem. We know that in statistical physics, entropy counts
the number of accessible microstates that a system can occupy, where
all states are presumed to occur with equal probability. On the other
hand, we also know that black holes can be completely characterized
by only three externally observable classical parameters: mass, electric
charge, and angular momentum. All other information about the matter
which formed a black hole ``disappears''
behind the black-hole event horizon, and therefore the nature of these
microstates is obscure. Then, what is the origin of the black hole
entropy? Furthermore, in order to justify the name ``entropy'',
one must to explain also why $S=S_{bh}+S_{out}$ is a non-decreasing
function of time, in other words, why black holes obey a ``generalized
second law of thermodynamics''.

We still have that, since black holes evaporate, one could expect,
from the Hawking radiation, any information about the state which
collapsed into the black hole. However, Hawking showed that this radiation
is thermal, and therefore does not carry any information about the
black hole initial state. That is to say, no information can escape
from inside of the black hole horizon. In this situation, the matter
that formed the black hole, which initially was in a pure state has
evolved into a mixed state. This bring us a contradiction with quantum
mechanics. There, a pure state can only evolve into another pure state
because of the unitarity of the evolution operator\cite{sw.hawking-cmp43,sw.hawking-prd14,j.preskill-hepth9209058,dn.page-hepth9305040}.

%


\indent One way to solve the Hawking paradox is that where information
can be stored in a topological disconnected region which arises inside
of the black hole \cite{sdh.hsu-plb644}. Gravitational collapse would
lead to a region of Planckian densities and curvature where quantum
gravitational effects can lead to a topology change process where
a new topologically disconnected region appears. However, This process had
been claim to break unitarity and cluster deposition(locality) \cite{sdh.hsu-plb644,l.susskind-hepth9405103}.

On the other hand, recently it was suggested by one of us that it is possible to realize a topology change process to black holes without break unitarity or
cluster decomposition \cite{c.a.s.silva}. The basic idea of this proposal is to see the black hole horizon as a fuzzy sphere $S_{F}^{2}$ taking into account some 
quantum symmetry properties related with
a Hopf algebra structure. There, we can define a linear operation
(the coproduct of the Hopf algebra) on $S_{F}^{2}$ and compose two fuzzy
spheres preserving algebraic properties intact. This operation, which
we shall represent by $\Delta$, produces a topology change process
where a fuzzy sphere splits in two others \cite{ap.balachandran-ijmpa19}.

Let M describes a wave function on $S_{F}^{2}$, the coproduct $\Delta:S_{F}^{2}(J)\rightarrow S_{F}^{2}(K)\otimes S_{F}^{2}(L)$
acts on M as
 \begin{eqnarray}
\Delta(M)_{(K,L)}&=&\hspace{-5mm}\sum_{\!\!\!\!\!\m_{1},\m_{2},m_{1},m_{2}}\hspace{-5mm}C_{K,L,J;\m_{1},\m_{2}}C_{K,L,J;m_{1},m_{2}} \nonumber \\
&\times& M_{\m_{1}+\m_{2},m_{1}+m_{2}}e^{\m_{1}m_{1}}(K)\otimes e^{\m_{2}m_{2}}(L)  \label{d-split}
\end{eqnarray}
where $C$'s are the Clebsh-Gordan coefficients and $e^{\m_{i}m_{j}}$'s
are basis for $Mat$. From \equ{d-split} M $\in$ $S_{F}^{2}(J)$
splits into a superposition of wavefunctions on $S_{F}^{2}(K)\otimes S_{F}^{2}(L)$,
and the information in M is divided between the two fuzzy spheres
with spins $K$ and $L$ respectively.

The process \equ{d-split} preserves the Hermitian conjugation, the matrix product, the trace, and the inner product.The two last properties assure that \equ{d-split} is a unitary
process \cite{ap.balachandran-ijmpa19}. 


\indent Following \cite{c.a.s.silva}, if we use the fuzzy sphere Hilbert space as the ones of
the black hole, we shall have the following consequences: 
(i) The maximum of information about the black hole that
an outside observer can obtain would be encoded in a wave function
M defined on the fuzzy sphere Hilbert space. (ii) We shall find out, through the Hopf coproduct $\Delta$
in \equ{d-split}, a topology change process for the black hole.
In this process the information about the black hole initial state,
described by the wave function $M$, is divided into two regions.
One of them is a fuzzy sphere with spin $K$, which we shall consider
as the original world and name it {}``the main world''. The other
one is a fuzzy sphere with spin $L$ which we shall name ``the
baby world''. (iii) Since an observer in the main world can not access
the degrees of freedom in the baby world, from his standpoint, the
black hole will appear to evolve from a pure to a mixed state, described
by a density matrix $\rho$. Therefore, we can define an entropy measured
by the observer in the main world.

The entropy measured by the observer in the main world would be 
given by $S$ = $-k_{B}\ln(\rho tr{\rho})$ = $k_{B}\ln(dimH)$, where $H$ is Hilbert space associated to the black hole. 
If we associate a Hilbert space $H_{i}$ with each cell on the fuzzy sphere, we shall
have that, for a fuzzy sphere with n cells, the hole Hilbert space
will be given by $H=\bigotimes_{i=1}^{n}H_{i}.$ Therefore, we shall have
$S = k_{B}\ln(dimH_{i})^{n}=nk_{B}\ln(dimH_{i})$. On the other hand, the fuzzy sphere area can be written as $A=\alpha nl_{P}^{2}$, 
where $\alpha$ is related with the quantization scale,
and $l_{P}^{2}$ is the Planck area.

In this way, we have that the  black hole entropy can be written as

\begin{equation}
S=\varepsilon k_{B}\frac{A}{l_{P}^{2}},\label{entropy-3}
\end{equation}
 
\noindent which corresponds to the Bekenstein-Hawking formula, unless
for the undetermined factor $\varepsilon=\ln(dimH_{i})/\alpha$. 

The quantum topology change model, in this way, shed some light on
the problem of the origin of the entropy associated to a black hole by an outside observer: it is generated because
of the non-unitary evolutions of the geometry of the main world. As we shall show, the non-unitary evolution of the black hole geometry can also be the
origin of the GSL.


Now let us address the black
hole evaporation process, as it is seen by an observer in our universe,
through the topology change \equ{d-split}. In the intend to do this, at first, let us address the black hole area spectrum. From the equation \equ{entropy-3},
this spectrum is given by

\begin{equation}
A_{j}=\epsilon^{-1}l_{p}^{2}\ln(2j+1)\ .\label{bh-area}\end{equation}

We can choose the splitting process \equ{d-split} in a way that
$j\rightarrow j-\frac{1}{2}$, in the main world, and from \equ{bh-area}
this it will result in a decrease of the black hole area, until when
$j=0$ it has completely evaporated. The logarithmic dependence of the black hole area spectrum on $j$
in the expression \equ{bh-area} tell us that, the decrease in the horizon area occurs in a continuous way to large values of $j$, and in a discrete way 
to small values of $j$, when the black hole approaches the Planck scale, as have been shown in the figure \ref{fig:2}:

\begin{figure}[htb]
\centering 
\fbox{\includegraphics[width=5.5cm,height=3.5cm]{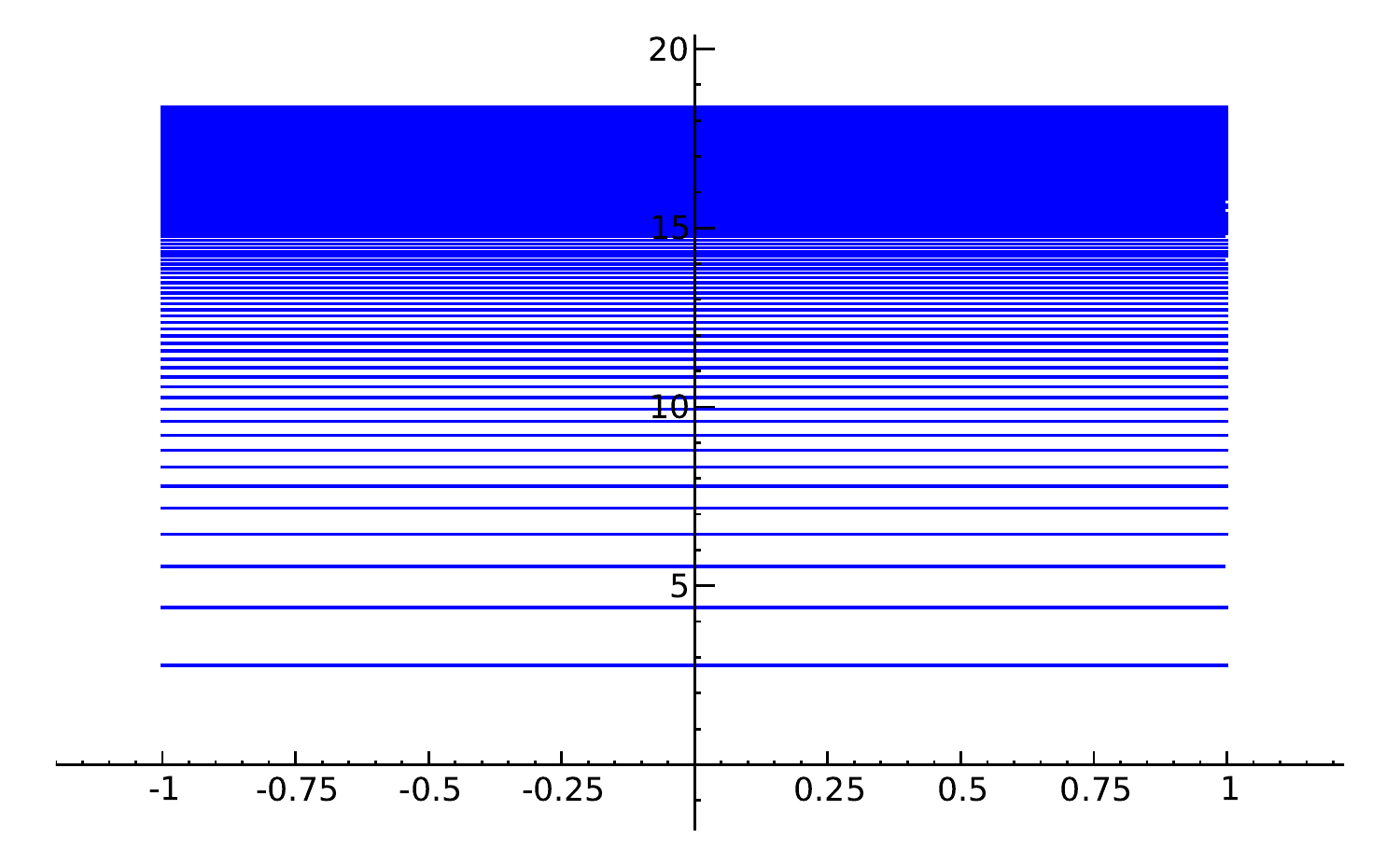}} 
\caption[Figure 2:]{Area spectrum, in units of Planck area for $\epsilon=1/4$, to a
quantum black hole in topology change approach. From the logarithmic dependence of the black hole area spectrum on $j$
in \equ{bh-area}, we have that the levels become
continuous for larger area values.}

\label{fig:2} 
\end{figure}

We have that the Hawking radiation is known semi-classically to be
continuous. However, the Hawking quanta of energy are not able to
hover at a fixed distance from the horizon since the geometry of the
horizon has to fluctuate, once quantum gravitational effects are included.
Thus, one suspects a modification of the black hole radiation, when
quantum geometrical effects are taken into account. Then any modification on the description
of the black hole emission process must occur at the final stages
of black hole evaporation, where its area spectrum becomes discrete.


To address it, let us analyze the process which consists in to go one step
down in the black hole area spectrum $(j\rightarrow j-1/2)$. From
equation \equ{d-split}, we have that, tracing over the degrees
of freedom in the baby universe, the splitting process \equ{d-split},
for a matrix $M=\hspace{2mm}\mid j,m\rangle\langle j,m^{'}\mid$ with
$l=1/2$, and $k=j-1/2$ is given by: 

\begin{eqnarray*}
\Delta(\mid j,m\rangle\langle j,m^{'}\mid)  & \hspace{-5cm} =  
\\  & \hspace{-10mm} \frac{\sqrt{(k+m_{k}+1)(k+m_{k}^{'}+1)}}{2k+1}\mid k, m-1/2\rangle\langle k, m^{'}-1/2\mid\\ 
 & \hspace{-12.5mm} +  \frac{\sqrt{(k-m_{k}+1)(k-m_{k}^{'}+1)}}{2k+1}\mid k, m+1/2\rangle\langle k, m^{'}+1/2\mid.\end{eqnarray*}

In this way, the probability to go down one step in the black hole area spectrum is given by

\begin{equation}
p_{j\rightarrow k\,=\, j-1/2}=phase\times\Big(\frac{2k+2}{2k+1}\Big) = phase\times\; e^{\frac{-\epsilon\delta A}{l^{2}_{p}}} \;,\label{prob-1}
\end{equation}

\noindent where a normalization factor $(2k+1)^{-1}$ has been included.

The probability for a black hole goes n steps down into its area spectrum
can be obtained from equation \equ{prob-1}. In reference \cite{c.a.s.silva}
it been shown that the splitting process \equ{d-split} obeys cluster
decomposition. In this way, we have that the different steps $j\rightarrow j-1/2$
in the black hole evaporation are independent events. Then the probability
of n steps occur in the black evaporation process is given by the
product of the probabilities of each one of this steps occurring by
itself. Then we have,

\begin{equation}
p_{kn}=phase\times\Big(\frac{2k+2}{2k-n+1}\Big)=phase\times\; e^{\frac{-\epsilon\delta A_{kn}}{l^{2}_{p}}},\label{prob-2}
\end{equation}

The probability above depends on the undetermined parameter $\epsilon$, which appears in the expression to the black hole entropy \equ{entropy-3}.
We have that, from the equation \equ{prob-2}, the density matrix describing the black hole is given by 
$\hat{\rho} \sim \sum_ {j} e^{-\epsilon A_ {j}} \mid j\rangle\langle j\mid = e^{-\epsilon \hat{A}}$. 
In this way, the black hole density matrix must satisfy the equation

\begin{equation}
i \frac{\partial \hat{\rho}}{\partial \Theta} = - \frac{\hat{A}}{8 \pi} \hat{\rho}  \label{evo-eq}
\end{equation}
 
As addressed by \cite{s.carlip-cqg12} and \cite{s. massar-nucphysb575}, the parameter 
$\Theta = 8 \pi i \epsilon$ plays the role of
a sort of ``dimensionless internal time'' associated with the horizon. The equation \equ{evo-eq} must be used when working in the Euclidean continuation 
of the black hole \cite{m.banados-prl72}. Regularity of the Euclidean manifold at the horizon imposes a fixed Euclidean angle given by $\Theta = 2 \pi i$ 
\cite{m.banados-prl72}, \cite{s.carlip-cqg12}, and \cite{s. massar-nucphysb575}. In this way, the undetermined parameter in \equ{entropy-3} and 
\equ{prob-2} is fixed as $ \epsilon = 1/4$.

The results above revel the evolution of the black hole geometry induced by the topology change process \equ{d-split} in the evaporation process, and bring
essential consequences for the way how entropy is emitted
in the black hole evaporation process. 

The entropy of a system measures
one's lack of information about its actual internal configuration
\cite{c.e.shannon-mtc,e.t.jaynes-pr106,e.t.jaynes.1-pr108,j.d.bekenstein-prd7}.
Suppose that all that is known about the internal configuration of
the system is that it may be found in any of a number of states, with
probability $p_{n}$ for the nth state. Then the entropy associated
with the system is given by Shannon's well-known relation $S=-\sum p_{n}lnp_{n}$.

The probability for a black hole to emit a specific quantum should
be given by the expression \equ{prob-2}, in which we shall include
a gray-body factor $\Gamma$ (representing a scattering of the quantum
off the spacetime curvature surround the black hole). Thus, the probability
$p_{n}$ to n steps in the mass (area) ladder is proportional to $\Gamma(n)\; e^{-\frac{\delta A_{kn}}{4l_{p}^{2}}}$.

The discrete mass (area) spectrum \equ{bh-area} implies a discrete
line emission from a quantum black hole. For a Schwarzschild black
hole, the radiation emitted by the black hole will be at frequencies
given by $\omega=\frac{1}{2\sqrt{\pi}}(\sqrt{ln(2j+1)}-\sqrt{ln(2j+1-n)})$.

To gain some insight into the physical problem, we shall consider
a simple toy model suggested by Hod \cite{S.Hod}. It is well known
that, for massless field, $\Gamma(M\omega)$ approaches $0$ in the
low-frequency limit, and has a high-frequency limit of $1$ \cite{dn.page-prd13, dn.page-prd14, dn.page-prd16}.That
is, $\Gamma(\bar{\omega})=0$ for $\bar{\omega}<\bar{\omega}_{c}$,
and $\Gamma(\bar{\omega})=1$ otherwise, where $\bar{\omega}=M\omega$.

The ratio $R=\mid\!\!\!\!\!\!\dot{\;\;\; S_{rad}}/\!\!\!\!\!\!\dot{\;\;\; S_{BH}\!}\mid$
of entropy emission rate from the quantum black hole to the rate of
black hole entropy decrease is given by:

\begin{equation}
R=\frac{-\sum_{i=1}^{N_{s}}\sum_{n=1}^{2k+1}C\Gamma(n)e^{-\frac{\delta A_{kn}}{4}}ln\Big[C\Gamma(n)e^{-\frac{\delta A_{kn}}{4}}\Big]}
{\sum_{i=1}^{N_{s}}\sum_{n=1}^{2k+1}C\Gamma(n)e^{-\frac{\delta A_{kn}}{4}}\Big(\frac{\delta A_{kn}}{4}\Big)}\;,\label{R-form}\end{equation}
 where C is a normalization factor, defined by the normalization condition:

\begin{equation}
\sum_{i=1}^{N_{s}}\sum_{n=1}^{2k+1}C\Gamma(n)e^{-\frac{\delta A_{kn}}{4}}=1\;\;.\label{norm.cond}\end{equation}

$N_{s}$ is the effective number of particle species emitted ($N_{s}$
takes into account the various modes emitted). We shall consider $N_{s}\propto2k+2$,
with a proportionality constant less than or equal to one, since the
modes emitted by the black hole in our treatment must be upper limited
by the number of degrees of freedom on the fuzzy sphere. 

We have plotted R down, taking $\bar{\omega}_{c}\simeq0.2$ (the location
of the peak in the total power spectrum \cite{dn.page-prd13, dn.page-prd14, dn.page-prd16}).

\begin{figure}[htb]
 \centering 
 \fbox{\includegraphics[width=5.5cm,height=3.5cm]{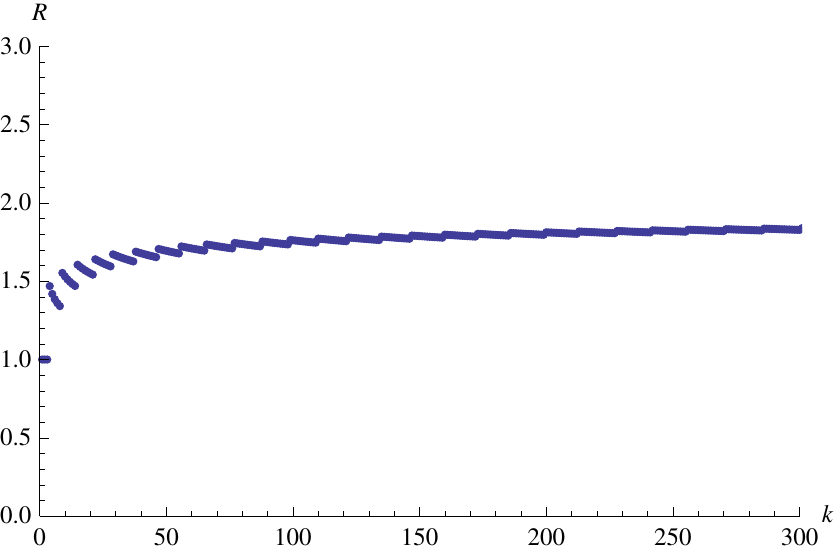}} 
 \label{fig:4} 
\end{figure}

From the graphic above, we have that the emission process respect
a ``second law of thermodynamics'', since
R is ever larger than (or equal) to unity. In this way, the non-unitary evolution of the black hole geometry in the main world, due to the topology change
process, can be the origin of the GSL.

Besides it is important to notice the entropy emitted from the black hole
decreases as the area spacing increases. In this way, the entropy of the radiation
emitted by the black hole becomes increasingly smaller with each step
of the evaporation process, mainly when the black hole reaches the
Planck scale ( notice that R reaches the unity in the last steps).


In this work we have argued that the quantum topology change model
to black hole evaporation, proposed by us in reference \cite{c.a.s.silva},
shed some light on the problem of the origin of black hole entropy:
it is generated because of the non-unitary evolutions of the geometry
of the main world. Besides, the topology change approach give us a
relation of states to points that agrees with our standard concept
of entropy as the logarithm of the number of microstates, from which
we have derived the Bekenstein-Hawking formula, $S = A/4$.

From the topology change model we have obtained a black hole area
spectrum, which is continuous in the classical/semiclassical limit,
and becomes discrete as the black hole approaches the Planck scale.
As a consequence of this, we have that information can leak out from
black hole, since its radiation becomes less and less entropic as
the black hole evaporates. In this way, some information about the black
hole initial state could be accessible to an observer in our universe.
Since it would occurs more strongly in the quantum gravity limit,
it does not require radical modifications in the laws of physics above
the Planck scale. The task of found an appropriate quantum mechanism
for information leak remains.

We have also shown that the description of the black hole evaporation
through a quantum topology change process proposed by \cite{c.a.s.silva}
could be the origin of the GSL. 



The authors thanks to Coordena\c{c}\~{a}o de Aperfeiçoamento de Pessoal de Ensino
Superior-CAPES(Brazil), and Conselho Nacional de Desenvolvimento Científico e Tecnológico - CNPQ(Brazil) for the financial support, and  C.A.S. Almeida 
for the careful reading of the manuscript


\expandafter\ifx\csname url\endcsname\relax \global\long\def\url#1{\texttt{#1}}
\fi \expandafter\ifx\csname urlprefix\endcsname\relax\global\long\def\urlprefix{URL }
\fi

\end{document}